
\centerline{\bf Axisymmetric Stationary Solutions as Harmonic Maps}
\vskip 1cm

\centerline {by}

\centerline {T. Matos and J. Pleba\'nski+}

\centerline {Departamento de Fisica, Centro de Investigacion y de Estudios
Avanzados }

\centerline {del IPN, Apdo. Post. 14-740, Mexico 07000 DF, Mexico}
\vskip 1cm

\noindent
Abstract: We present a method for generating exact solutions of Einstein
equations in vacuum using harmonic maps, when the spacetime possesses two
commuting Killing vectors.  This method consists in writing
the axisymmetric stationary
Einstein equations in vacuum as a harmonic map which belongs to the group
$SL(2,{\bf R})$, and decomposing it in its harmonic $``$ submaps$''.$ This
method provides a natural classification of the solutions in classes
(Weyl's class, Lewis' class etc.).\hfil\break
\vskip 1cm
\noindent Pacs: 04.20-Cv, 04.20.Jb
\vskip 1.5cm
\noindent + On leave of absence of Warsaw University
\vfil\eject
\noindent
Introduction\hfil\break
\indent
Stationary Einstein equations in terms of harmonic maps (HM) [1] are
well-known.  S\'anchez [2] found that the stationary Einstein equations in
electrovacuum are in fact HM, but that the corresponding differential
equations are non-linear.  Neugebauer and Kramer [3] showed that this HM
has a Lax pairs-representation (equivalent linear problem) in terms of a
HM which is an element of the group SU(2,1), i.e. axisymmetric stationary
Einstein-Maxwell equations in potential space (Ernst potential space) are
HM $G:M^{4}\rightarrow V_p\otimes SU(2,1)$, being $V_p$ a p-dimensional
Riemannian space. On the other hand, Belinsky and Zakharov [4] have
also given the empty Einstein equations in terms of HM, but in a
$SL(2,{\bf R})$ representation of spacetime.  Recently, Whitman and
Stoeger gave a HM representation of these equations [5]. It is also
possible to give the empty Einstein equation in potential space in terms
of HM if we reduce the group SU(2,1) to SU(1,1).  The isomorphism
$SU(1,1) \simeq SL(2,{\bf R})$ allows us to give a representation of
these equations in terms of the group $SL(2,{\bf R})$ [6].
Thus we can write the Einstein equations in vacuum as a
$SL(2,{\bf R})$ -HM in the spacetime,
and in the potential space.\hfil\break
\indent
In this work, we will start from the HM representation of the axisymmetric
stationary Einstein equations in vacuum, and develop a method for generating
exact solutions.  It consists in separating the group $SL(2,{\bf R})$
into its one-parametric
subgroups and itself giving a representation of the HM, $G:M^{4}
\rightarrow V_p\otimes SL(2,{\bf R})$ in its tangent space (see section 3 or
ref. [7]).  The result is that each subgroup corresponds to a well-known
class of solutions, explaining why do these classes exist.\hfil\break
\indent
One could also extend this analysis to the Einstein-Maxwell equations.
It may be done in potential space [8] where the HM belongs to the group
$SU(2,1)$, but there is no HM representation in spacetime. Nevertheless,
five-dimensional gravity admits a HM re\ presentation for the stationary
equations [9].  Both the spacetime and the potential space have a
$SL(3,{\bf R})$ HM representation of the field equations [10]
(see also ref. [9]).  Thus an analysis with harmonic $``$ submaps$''$
can be done [11],[12].\hfil\break
\indent
This work pretends to be self-contained.  In sections 1 and 2 we give the
Einstein's spacetime field equations in vacuum when the metric depends only
on two coordinates. In section 3 we briefly outline the stationary Einstein
equations in potential space (Ernst potential).  Section 4 is
devoted to explain the method of subgroups (harmonic $``$ submaps $''$ ) which
will be applied in section 5. \hfil\break
\hfil\break
\noindent
{\bf 1. Generalities}\hfil\break
\indent
We consider a $V_4$ manifold with two commuting Killing-vectors and
local coordinates $\{X^{\mu}\}=(X^r,y^A)$,
with $X^{r}=(X^{1},X^{2})$ being the $``$ active $''$ (and
arbitrary) coordinates and $y^{A}$ = $(y^{1}, y^{2})$ playing the role
of the $``$ passive $''$ coordinates (i.e., with
$\partial /\partial y^{A}$ being the two commuting Killing vectors).
The metric can be given in the quasi-diagonalizable form \hfil\break
$$V_{4}: g_{4}=g_{rs}dX^{r}\otimes_{_{_{_{\kern -2.1mm {\tenbf s}}}}}
        dX^{s}+K_{AB}dy^{A}\otimes_{_{_{_{\kern -2.1mm {\tenbf s}}}}}
dy^{B}.\eqno(1.1)$$
\noindent
The essential $``$Killigian$''$ part of the metric is thus described by the
three real objects\hfil\break
$$K_{AB}=K_{AB}(X^{1},X^{2})=K_{(AB)}.\eqno(1.2)$$
\noindent
We use for this part of the metric the spinorial notation,
because evidently the $constant$\hfil\break
$$GL(2,{\bf R})\eqno(1.3)$$
\noindent
transformations of the passive variables $y^{A}$ do not affect the structure
considered in any manner.\hfil\break
\indent
We assume the rules of manipulating the spinorial indices as
follows:\hfil\break
$$\psi_{A} = \epsilon_{AB}\psi^{B}\Leftrightarrow \psi^{A}=
\psi_{B}\epsilon^{BA}$$
$$\Vert\epsilon_{AB}\Vert=\left( \matrix{0 & 1\cr
                                      -1 & 0\cr}\right)
=\Vert\epsilon^{AB}\Vert .\eqno(1.4)$$
\indent
We assume the $``$ Killingian sector $''$ to be of signature (+,$-$). To ensure
this, it is necessary and
sufficent to assume that\hfil\break
$$K:=det(K_{AB}) = {1\over{2}}K_{AB}K^{AB}< 0.\eqno(1.5)$$
\indent
The 2-dimensional metric\hfil\break
$$V_{2}: \ \ \ g_{2}:=g_{rs} dX^{r}\otimes dX^{s}\eqno(1.6)$$
\noindent
with $g_{rs}=g_{rs}(X^{1},X^{2})=g_{(rs)}$ has then the signature
(+,+). The condition for this is\hfil\break
$$det (g_{rs})>0,\eqno(1.7)$$
\noindent
assuming, of course, that the coordinates $X^{r}$ are real.  The
$V_{2}$ structure is thus just a proper 2-dimensional Riemannian space.  The
covariant derivative with respect to $g_{2}$ shall be denoted by
$``|r''$. The indices $r,s,...$ are to be manipulated by $g_{rs}$ and its
inverse, the contravariant metric $g^{rs} \ \ \
(g^{rs}g_{st}=\delta^{r}_{t})$.\hfil\break
\indent
Our problem consists now in studying the Einstein empty spacetime equations
working with the chart $\{X^{\mu}\}=
\{X^{r},y^{A}\},$ \hfil\break
$$G_{\mu\nu}=0,\eqno(1.8)$$
\noindent
which we can evidently also take in the form \hfil\break
$$R_{\mu\nu}=0,$$
\noindent
with $R_{\mu\nu}={R^{\sigma}}_{\mu\nu\sigma}$ being the Ricci
tensor.\hfil\break
\indent
The Ricci tensor of metric (1.1) can be easily evaluated in the
covariant form with respect to the chart used in the description of the
metric $g_{2}$ ,i.e., the coordinates $\{X^{r}\}=
\{X^{1},X^{2}\}.$\hfil\break
\indent
Indeed, one finds that all $``$ mixed $''$ components of
$R_{\mu\nu}$ identically vanish \hfil\break
$$R_{Ar} =0= R_{rA}.\eqno(1.9)$$
\indent
For the $``$ Killingian $''$ components of $R_{\mu\nu}$ one finds
the $V_{2}$ -covariant basic expression \hfil\break
$$R_{AB} ={1\over{2}}K_{AB|r}^{\ \ \ \ |r}+{1\over{4}}K^{-1}K^{|r}
K_{AB|r}-{1\over{2}}K^{-1}K^{PQ}K_{PA|r}K_{QB}^{\ \ \ |r},\eqno(1.10)$$
\noindent
with K defined by (1.5) and all spinorial indices manipulated according to
the rules (1.4).\hfil\break
\indent
Now, the curvature of $V_{2}$ is of course characterized entirely by its scalar
curvature. If ${\buildrel 0\over R}{}^{r}_{\kern 1mm stu}$ is the curvature
tensor of $V_{2}$ , we have \hfil\break
$${\buildrel 0\over R}{}^{rs}_{\kern 3mm tu}=- {1\over{2}}
\delta^{rs}_{tu}{\buildrel 0\over R},\eqno(1.11)$$
\noindent
with ${\buildrel 0\over R}$ being the scalar curvature of the non-Killingian
sector $(g_{2})$ of the metric.\hfil\break
\indent
Knowing this, one works out the $``$ non-Killingian$''$ components
of $R_{\mu\nu}$ ,i.e. $R_{rs}$ , in the form \hfil\break
$$R_{rs}={1\over{2}}g_{rs}{\buildrel 0\over R}+ {1\over{2}}
(K^{-1}K_{,r})_{|s}+{1\over{4}}K^{-2}K^{PU}K_{QU ,r}K^{QV}
K_{PV,s}.\eqno(1.12)$$
\indent
The next thing to do is to work out the components of $R_{AB}$ and $R_{rs}$
in an apropiate form for our further purposes.\hfil\break
\indent
For $R_{AB}$ we notice first that we have the
spinorial identity\hfil\break
$$K^{PQ}K_{PA}=K\delta^{Q}_{A}.\eqno(1.13)$$
\noindent
Employing it, we find that\hfil\break
$$K^{PQ}K_{PA\vert r}K_{QB}^{\ \ \ |r} $$
$$ ={(K_{|r}\delta}^{{Q}}_{A}-K^{PQ}_{\ \ \ |r}K_{PA})
K_{QB}^{\ \ \  |r} \eqno(1.14)$$
$$=K^{|r}K_{AB|r}-K_{AP}{K^{QP}}_{|r} K_{QB}^{\ \ \ |r}$$
\noindent
but
$K^{Q}_{\ P|r}K_{QB}^{\ \ \ |r}=- K^{Q}_{\ B|r}K_{QP}^{\ \ \ |r}$
implies that\hfil\break
${1\over {2}}\epsilon_{PB}\epsilon ^{P'B'}{K^{Q}}_{P'|r} K_{QB'}^{\ \ \ |r}=
{1\over{2}}\epsilon_{PB}K^{MN|r}K_{MN|r}$. Equality (1.14) says then
that\hfil\break
$$K^{PQ}K_{PA|r}K_{QB}^{\ \ \ |r}=K^{|r}K_{AB|r}+K_{A}^{\ P}{K^Q}_{P|r}
K_{QB}^{\ \ \ |r}$$
$$=K^{|r}K_{AB|r}+{1\over{2}}K_{A}^{\ P}\epsilon_{PB}K^{MN|r}
K_{MN|r}.\eqno(1.15)$$
\indent
We arrive in this way at the identity\hfil\break
$$K^{PQ}K_{PA|r}K_{QB}^{\ \ \ |r}=K^{|r}K_{AB|r}-{1\over{2}}K_{AB}
K^{MN|r}K_{MN|r}.\eqno(1.16)$$
\noindent
Using this in (1.10) we have\hfil\break
$$R_{AB}={1\over{2}}{K_{AB|r}}^{|r}+{1\over{4}}K^{-1}K_{AB|r}$$
$$-{1\over{2}}K^{-1}K^{|r}K_{AB|r}+{1\over{4}}K^{-1}K_{AB}K^{MN|r}
K_{MN|r}$$
$$={1\over{2}}K_{AB|r}^{\ \ \ \ |r}-{1\over{4}}K^{-1}K^{|r}K_{AB|r}+
{1\over{4}}K^{-1}K_{AB}K^{MN|r}K_{MN|r}.\eqno(1.17)$$
\indent
We bring this last expression to the slightly simpler form \hfil\break
$$R_{AB}={1\over{2}}\sqrt{-K}({1\over{\sqrt{-K}}}K_{AB|r})^{|r}+
{1\over{4}}K^{-1}K^{MN|r}K_{MN|r}.$$
\noindent
The expression above is already satisfactory for many purposes.
We can obtain a more satisfactory equivalent expression by executing
the contraction $K^{AP}R_{AB}$, remembering that because of $det(K_{AB})<0$,
the matrix $K_{AB}$ is invertible.
Indeed, employing (1.13) in the form
$K^{AP}K_{AB}=K\delta^{P}_B$ , we have\hfil\break
$$K^{AP}R_{AB}={{{1\over{2}}\sqrt{-K}}}K^{AP}
({1\over{\sqrt{-K}}}K_{AB|r})^{|r}+
{1\over{4}}\delta^{P}_B K^{MN|r}K_{MN|r} $$
$$={1\over{2}}\sqrt{-K}\{({1\over{\sqrt{-K}}}K^{AP}K_{AB|r})^{|r}-
K^{AP}_{\ \ \ |r}{1\over{\sqrt{-K}}}K_{AB}^{\ \ \ |r}\}$$
$$+{1\over{4}}\delta^{P}_{B}K^{MN|r}K_{MN|r} $$
$$={1\over{2}}\sqrt{-K}({1\over{\sqrt{-K}}}K^{AP}K_{AB|r})^{|r}+$$
$${1\over{4}}\delta^{P}_{B}K^{MN|r}K_{MN|r}-{1\over{2}}K^{AP}_{\ \ \ |r}
K_{AB}^{\ \ \ |r}.\eqno(1.19)$$
\noindent
The last term in this expression, with the index P lowered,
$-{1\over{2}}K^{A}_{\ P|r}K_{AB}^{\ \ \ |r}={1\over{2}}K^{A}_{\ B|r}
K_{AP}^{\ \ \ |r}$ is
antysymmetric in $PB$ and hence equal to\hfil\break
$$-{1\over{2}}{K^A}_{P|r}K_{AB}^{\ \ \ |r}=-{1\over{4}}\epsilon_{PB}
\epsilon^{P'B'}{K^A}_{P'|r} K_{AB}^{\ \ \ |r}\eqno(1.20)$$
$$=-{1\over{4}}\epsilon_{PB}K^{MN|r}K_{MN|r},$$
\noindent
so that, rising again the index $P$\hfil\break
$$-{1\over{2}}K^{AP}_{\ \ \ |r}K_{AB}^{\ \ \ |r}=
-{1\over{4}}\delta^{P}_{B}K^{MN|r}K_{MN|r}.\eqno(1.21)$$
\noindent
Consequently, the terms in the last line of (1.19) simply cancel out, and
we arrive at \hfil\break
$$E_{AB}:={K_A}^{R}R_{RB}\equiv {1\over{2}}{\sqrt{-K}}
({{1}\over{\sqrt{-K}}}{K_A}^{R}K_{RB|r})^{|r}.\eqno(1.22)$$
\indent
This is perhaps the most condensed form concerning the analytic form
of the $``$ Killingian part $''$ of the Ricci tensor.  We would like to
observe at this point that the expression $E_{AB}$ defined
above has the trace \hfil\break
$${E^A}_{A}\equiv K^{AB}R_{AB} $$
$$={1\over{2}}\sqrt{-K}({{1}\over{\sqrt{-K}}}K^{AB}K_{AB|r})^{|r}=
{1\over{2}}\sqrt{-K}({{1}\over{\sqrt{-K}}}K_{|r})^{|r} $$
$$=-{1\over{2}}\sqrt{-K}[{1\over{\sqrt{-K}}}(\sqrt{-K})^{2}_{\ |r}]^{|r} $$
$$=-\sqrt{-K}[(\sqrt{-K})_{|r}]^{|r}.\eqno(1.23)$$
\noindent
Moreover\hfil\break
$$K^{AB}E_{AB}=K^{AB}K_{A}^{\ R}R_{RB}=K\epsilon^{BR}R_{RB}\equiv
0,\eqno(1.24)$$
\noindent
so that the objects $E_{AB}$ defined by (1.22) are
$\underline {linearly \ dependent.}$\hfil\break
\indent
In the next part of this section, we would like to study the structure
of the $``$ non-Killingian$''$ part of the Ricci tensor, (1.2).\hfil\break
\indent
Because of (1.13), we have\hfil\break
$$K^{UP} K_{UQ|r}=- K_{UQ}K^{UP}_{\ \ \ |r}+
K_{|r}\delta^{P}_{Q}\eqno(1.25)$$
\noindent
and therefor the last term in (1.12) transforms to
$${1\over{4}}K^{-2}K^{UP}K_{UQ|r}K^{QV}K_{PV|s}$$
$$={1\over{4}}K^{-2}(K_{|r}\delta^{P}_{Q}-K_{UQ}K^{UP}_{\ \ \ |r})
K^{QV}K_{PV|s}$$
$$={1\over{4}}K^{-2}K_{|r}K^{PV}K_{PV|s}-{1\over{4}}K^{-2}
K^{QV}K_{QU} K^{UP}_{\ \ \ |r}K_{VP|s}$$
$$={1\over{4}}K^{-2}K_{|r}K_{|s}-{1\over{4}} K^{-2}
K \delta^{V}_{U}K^{UP}_{\ \ \ |r}K_{VP|s}$$
$$={1\over{4}}K^{-2}K_{|r}K_{|s}-{1\over{4}}K^{-1}
K^{MN}_{\ \ \ |r}K_{MN|s}.\eqno(1.26)$$
\indent
With this identity, we have\hfil\break
$$R_{rs}={1\over{2}}g_{rs}{\buildrel 0\over R} +{1\over{2}}
(K^{-1}K_{,s})_{|s}+{1\over{4}}K^{-2}K_{|r}K_{|s}
-{1\over{4}}K^{-1}K^{MN}_{\ \ \ |r} K_{MN|s}.\eqno(1.27)$$
\indent
This can be still slightly simplified, observing that\hfil\break
$${1\over{2}}(K^{-1}K_{,s})_{|s}+{1\over{4}}K^{-2}K_{|r}K_{|s}\equiv
{1\over{\sqrt{-K}}}(\sqrt{-K})_{|rs}.\eqno(1.28)$$
\indent
Hence, we have for $R_{rs}$\hfil\break
$$R_{rs}={1\over{2}}g_{rs}{\buildrel 0\over R} + {1\over{\sqrt{-K}}}
(\sqrt{-K})_{|rs}
-{1\over{4}}K^{-1}K^{MN}_{\ \ \ |r}K_{MN|s}.\eqno(1.29)$$
\indent
We can now conveniently evaluate the scalar curvature of the $g_{4}$ metric.
Indeed, it is obvious from (1.1) that\hfil\break
$$ \Vert g_{\mu \nu} \Vert =\left(
\vbox{\offinterlineskip\tabskip=3pt
 	\halign{\strut
      		\hfil#\hfil &\vrule# &\hfil#\hfil\cr
      		$g_{rs}$ & &0\cr
      		\noalign{\hrule}
      		0 & & $K_{AB}$ \cr}}\right)\ ,\ \ \
\Vert g^{\mu \nu}\Vert=\left(
\vbox{\offinterlineskip\tabskip=3pt
        \halign{\strut
                \hfil#\hfil &\vrule# &\hfil#\hfil\cr
                $g^{rs}$ & &0\cr
                \noalign{\hrule}
                0 & & ${K^{AB}\over{K}}$ \cr}}\right)\ ,$$
$$ \eqno(1.30) $$
\noindent
and consequently\hfil\break
$$R=g^{rs}R_{rs}+{1\over{K}}K^{AB}R_{AB},$$
\noindent
remembering (1.9).  From (1.29) we evaluate now\hfil\break
$$g^{rs}R_{rs}={\buildrel 0\over R}{1\over{\sqrt{-K}}}
(\sqrt{-K})_{|r}^{\ |r}-{1\over{4}}K^{-1}K^{MN|r}K_{MN|r} \eqno(1.32) $$
\noindent
and from (1.23) we have\hfil\break
$$ {1\over{K}}K^{AB}R_{AB}=-{\sqrt{-K}\over{K}}(\sqrt{-K})_{|r}^{\ |r}=
{1\over{\sqrt{-K}}}(\sqrt{-K})_{|r}^{\ |r} \eqno(1.33) $$
\noindent
so that\hfil\break
$$R={\buildrel 0\over R}+{2\over{\sqrt{-K}}}(\sqrt{-K})_{|r}^{\ |r}-
{1\over{4}}K^{-1}K^{MN|r}K_{MN|r} .\eqno(1.34) $$
\indent
Knowing R, we can now evaluate the components of the Einstein tensor\hfil\break
$$G_{\mu\nu}:=R_{\mu\nu}-{1\over{2}}g_{\mu\nu}R.\eqno(1.35)$$
\noindent
In particular, the $(rs)$ components (i.e., the non-Killingian part) of this
object are\hfil\break
$$G_{rs}=R_{rs}-{1\over{2}}g_{rs}R $$
$$={1\over{2}}g_{rs}{\buildrel 0\over R}+{1\over{\sqrt{-K}}}(\sqrt{-K})_{|rs}
-{1\over{4}}K^{-1}K^{MN}_{\ \ \ |r}K_{MN|s} $$
$$-{1\over{2}}g_{rs}\{ {\buildrel 0\over R}+{2\over \sqrt{-K}}
(\sqrt{-K})_{|t}^{\ \ |t}-
{1\over{4}}K^{-1}{K^{MN}}_{|t}{K_{MN}}^{|t} \}$$
$$={1\over{\sqrt{-K}}}(\sqrt{-K})_{|rs}-{1\over{4}}K^{-1}
{{K^{MN}}_{|r}}K_{MN|s} $$
$$-g_{rs}\{{1\over{\sqrt{-K}}}{(\sqrt{-K})_{|t}}^{|t}-{1\over{8}}
K^{-1}{K^{MN}}_{|t}{K_{MN}}^{|t}\}. \eqno(1.36)$$
\indent
The cancellation of the term which involves ${\buildrel 0\over R}$ in this
expression is of basic importance, the second derivatives of the $g_{2}$
metric $({g_{rs,tu}})\ {do\ not}$ enter in the structure of
$G_{rs}$ .\hfil\break
\indent
Summarizing, we can now state the basic equations of the problem
$R_{\mu\nu}=0$ written covariantly with respect to the chart
$\{X^{r}\}=\{X^{1},X^{2}\}$ in terms of which the metric $g_{2}$ is
described as the following hierarchy of differential equations. First we
have the $``$ K-sector$''$ Einstein equations\hfil\break
$$\{det(K_{AB})\not=0\}\Rightarrow\{R_{AB}=0\Leftrightarrow E_{AB}:=
{K_{A}}^{R}R_{RB}=0\}, $$
\noindent
where\hfil\break
$$E_{AB}\equiv {1\over{2}}{\sqrt{-K}}\left({1\over{\sqrt{-K}}}
{K_A}^{R}{K_{RB}}_{|r}\right)^{|r}=0.\eqno(1.37)$$
\noindent
These equations imply\hfil\break
$${(\sqrt{-K})_{|r}}^{|r}=0,\eqno(1.38)$$
\noindent
so they are not independent because of the identity\hfil\break
$$K^{AB}E_{AB}\equiv 0.\eqno(1.39)$$
\noindent
Then we have the equations
$$G_{rs}\equiv {1\over{\sqrt{-K}}}({-K})_{|rs}-{1\over{4}}
K^{-1}{K^{MN}}_{|r}K_{MN|s}$$
$$-g_{rs}\{{1\over{\sqrt{-K}}}{(\sqrt{-K})_{|t}}^{|t}-{1\over{8}}
K^{-1}K^{MN|t}K_{MN|t}\}=0,\eqno(1.40)$$
\noindent
and finally\hfil\break
$$R\equiv {\buildrel 0\over R}+{2\over{\sqrt{K}}}{(\sqrt{-K})_{|r}}^{|r}
-{1\over{4}}K^{-1}K^{MN|r}K_{MN|r}=0,\eqno(1.41)$$
\noindent
where $\buildrel 0\over R$ is the scalar curvature of $g_{rs}$ defined
by (1.11).\hfil\break
\hfil\break
\indent
{\bf 2. The \ Equations \ $R_{\mu\nu}=0 $ \ in \ Weyl's
\ Coordinates.}\hfil\break
\noindent
The metric $g_{2}$ can be of course always expressed in the conformally
flat form\hfil\break
$$ g_{2}=\phi^{-2}\{ dX^{1}
                 \otimes_{_{\kern -2.1mm s}}
                               dX^{1} + dX^{2}
                               \otimes_{_{_{_{\kern -2.1mm {\tenbf s}}}}}
                               dX^{2}\},\eqno(2.1)$$
\noindent
or, introducing the complex coordinates\hfil\break
$$\xi :={1\over{\sqrt{2}}}(X^{1}+iX^{2})$$
$${\bar {\xi}}:={1\over{\sqrt{2}}}(X^{1}-iX^{2})\eqno(2.2)$$
\noindent
in the simple form \hfil\break
$$g_{2}=2\phi^{-2}d\xi
                        \otimes_{_{_{_{\kern -2.1mm {\tenbf s}}}}}
                        d{\bar{\xi}}\eqno(2.3)$$
\indent
We shall call $\xi$ and $\bar{\xi}$ the Weyl coordinates. They are
a sort of $``$ null variables$''$ and they are arbitrary
up to the transformations\hfil\break
$$\xi = f(\xi '),\eqno(2.4)$$
\noindent
with $f(z)$ being an arbitrary analytic function such that $f'(z)\not= 0.$
Under the transformation $\xi\rightarrow {\xi '}$ the real structural
function $\phi$ transforms according to\hfil\break
$$\xi\rightarrow {\xi '} \ \ , \phi\rightarrow
{\phi '} = \phi \vert f'(\xi)\vert^{-1}.\eqno(2.5)$$
\indent
Experience with the theory of the exact solutions, e.g., the cases
of the D-type solutions or the Tomimatsu-Sato solutions, indicates that the
Weyl variables are certainly not the best variables in practice, i.e., in
the description of the physically pertinent solutions.  On the other hand,
these variables are theoretically important in decoding the essential
structure of the differential problem stated at the end of the previous
section.\hfil\break
\indent
When we employ the Weyl coordinates, a simple manner of proceeding
(and taking advantage of the covariant formulation of the problem
outlined at the end of the previous section) is this: we simply
take the metric $g_{2}$ as given in the form\hfil\break
$$ g_{2}= 2\phi^{-2}dX^{1}
                              \otimes_{_{_{_{\kern -2.1mm {\tenbf s}}}}}
                              dX^{2}\eqno(2.6)$$
\noindent
understanding $X^{1}\equiv\xi$ and $\ X^{2}\equiv{\bar{\xi}}$. With these
complex coordinates all the apparatus of the classical
differential geometry in two dimensions, formally implicates the
condition on $g_{2}$ to be positive definite. This implys
that the determinant of the metric\hfil\break
$$\Vert g_{rs}\Vert = \phi^{-2}\left(\matrix{0&1\cr
                                            1&0\cr}\right)\eqno(2.7)$$
\noindent
must be negative,\hfil\break
$$det(g_{rs})=-\phi^{-4},$$
\noindent
so that\hfil\break
$$\sqrt{-det(g_{rs})}=\phi^{-2}.\eqno(2.9)$$
\indent
For the inverse metric we have then\hfil\break
$$\Vert g^{rs} \Vert = \phi^{2}\left( \matrix{ 0 & 1 \cr
                                             1 & 0 \cr }\right),\eqno(2.10)$$
\noindent
so that the tensor density\hfil\break
$${\cal G}^{tu}:={\sqrt{-det(g_{rs})}}g^{tu}\eqno(2.11)$$
\noindent
is just a numeric matrix\hfil\break
$$\Vert{\cal G}^{rs}\Vert =\left(\matrix{0&1\cr
                                        1&0\cr}\right) .$$
$$\eqno(2.12)$$
\indent
The Weyl coordinates are thus harmonic in the sense that\hfil\break
$${{\cal G}^{rs}}_{,s}=0.\eqno(2.13)$$
\indent
The Christoffel symbols $\Gamma^r_{\ st}$
computed from the metric (2.7) and its inverse (2.10)
vanish, excepting the two components\hfil\break
$$\Gamma^1_{\ 11}=-2({\it ln}\phi )_{,1}\equiv -2 ({\it ln}\phi)_{,\xi},$$
$$\Gamma^2_{\ 22}=-2({\it ln}\phi )_{,2}\equiv -2
({\it ln}\phi)_{,{\bar\xi}}.\eqno(2.14)$$
\indent
This permits us to determine easily the scalar curvature of $g_{2},$ the
object ${\buildrel 0\over R}$. We arrive at\hfil\break
$${\buildrel 0\over R}=g^{rs}{\buildrel 0\over R}{}^{t}_{\kern 1mm rst}=
-4\phi^{2}({\it ln}\phi)_{,\xi\bar \xi}.\eqno(2.15)$$
\indent
Now, equations of the type $(AB_{|r})^{|r}=0$ in the present coordinates,
being equivalent (when A and B are $g_{2}$ scalars) to
$({\cal G}^{rs}AB_{,r})_{,s}=0$, simply transform to
$(AB_{,\xi})_{,\bar\xi}+(AB_{,\bar\xi})_{,\xi}=0$ . Therefor, the fundamental
equations (1.37) take the form \hfil\break
$$R_{AB}=0\Leftrightarrow
({1\over \sqrt{-K}}{K_A}^R K_{RB,\xi})_{,\bar{\xi}} +
({1\over \sqrt{-K}}{K_A}^R K_{RB,\bar{\xi}})_{,\xi}=0.\eqno(2.16)$$
\indent
The essential fact emerges here since these
equations do not involve in any manner
the structural function $\phi$, being them the ${\underline autonomous}$
$``$ K-sector$''$ equations.  The necessary implication of these
equations (1.38), can be now simply stated in the form \hfil\break
$$(\sqrt{-K})_{,\xi\bar\xi}=0.\eqno(2.17)$$
\indent
Consider now equations (1.40), their (11) component transforms to\hfil\break
$$G_{11}={1\over{\sqrt{-K}}}(\sqrt{-K})_{|11}-{1\over{4}}K^{-1}
{K^{MN}}_{,1} K_{MN,1}=0,\eqno(2.18)$$
\noindent
or explicitly\hfil\break
$$\sqrt{-K}G_{11}=(\sqrt{-K})_{,\xi\xi}+2({\it ln}\phi)_{,\xi}
(\sqrt{-K})_{,\xi} + {1\over{4}}{1\over{\sqrt{-K}}}{K^{MN}}_{,\xi}
K_{MN,\xi}=0.\eqno(2.19)$$
\noindent
The complex conjugate of this equation transforms of course to\hfil\break
$$\sqrt{-K}G_{22}=(\sqrt{-K})_{,\bar\xi\bar\xi}+2({\it ln}\phi)_{,\bar\xi}
(\sqrt{-K})_{,\bar\xi}+{1\over{4}}{1\over{\sqrt{-K}}}{K^{MN}}_{,\bar\xi}
K_{MN,\bar\xi}=0.\eqno(2.20)$$
\indent
The $(12)$ component of equation (1.40), reads\hfil\break
$$G_{12}={1\over{\sqrt{-K}}}(\sqrt{-K})_{|12}-{1\over{4}}K^{-1}
{K^{MN}}_{,1} K_{MN,2}\eqno(2.21)$$
$$-\phi^{-2}\{{1\over{\sqrt{-K}}}{(\sqrt{-K})_{|r}}^{|r}-{1\over{8}}
K^{-1}2\phi^{2}{K^{MN}}_{,1}K_{MN,2}\}.\eqno(2.21)$$
\indent
But because of $\Gamma^r_{\ 12}\equiv 0$ and $(\sqrt{-K})_{|12}=
(\sqrt{-K})_{,\xi\bar\xi}$,
cancelling the terms with first derivatives, we have\hfil\break
$$G_{12}={1\over{\sqrt{-K}}}(\sqrt{-K})_{,\xi{\bar\xi}}-
{2\over{\sqrt{-K}}}(\sqrt{-K})_{,\xi{\bar\xi}} $$
$$=-{1\over{\sqrt{-K}}}(\sqrt{-K})_{,\xi{\bar\xi}}.\eqno(2.22)$$
\noindent
Including that
$R_{AB}=0$ implies $(\sqrt{-K})_{,\xi{\bar\xi}}=0$, it follows that
\hfil\break
$$R_{AB}=0\Rightarrow G_{12}=0,\eqno(2.23)$$
\noindent
so that the conditions $G_{12}=0$ are automatically fulfilled on the
$``$ K-sector$''$ equations.\hfil\break
\indent
The last of the field equations which must be described in terms of the
Weyl coordiantes is (1.41). Assuming that as a consequence of
$R_{AB}=0$ we know already that ${\sqrt{-K}_{|r}}^{|r}=0\Leftrightarrow
(\sqrt{-K})_{,\xi\bar\xi}=0$ , and using (2.15), this equation transforms to
\hfil\break
$$R=-4\phi^{2}({\it ln}\phi)_{,\xi\bar\xi}-{1\over{2}}
K^{-1}\phi^{2}{K^{AB}}_{,\xi}K_{AB,\bar\xi}=0,\eqno(2.24)$$
\noindent
or simply to demand that\hfil\break
$$ 8({\it ln}\phi)_{\xi{\bar\xi}}+{K^{-1}}{K^{AB}}_{{,\xi}}{K_{AB,{\bar\xi}}}
=0\eqno(2.25)$$
\indent
Summarizing, we conclude that working in the Weyl coordinates
$(\xi,{\bar\xi})$ , in order to fulfill the field equations
$R_{\mu\nu}=0$ , we must demand that the real structural functions
$K_{AB}=K_{AB}(\xi,{\bar\xi})=K_{(AB)},\phi =\phi(\xi ,{\bar\xi})$ should
be subject to the sequence of differential conditions:\hfil\break
\noindent
(a) $$ ({1\over{\sqrt{-K}}}{K_A}^RK_{RB,\xi})_{,\bar\xi}+
({1\over{\sqrt{-K}}}{K_A}^RK_{RB,{\bar\xi}})_{,\xi} =0\eqno(2.26) $$
\noindent
(b)$$\cases{ (\sqrt{-K})_{,\xi\xi} + 2({\it ln}\phi)_{,\xi}(\sqrt{-K})_{,\xi}
 + {1\over{4} }{1\over{\sqrt{-K}} }{K^{AB} }_{,\xi} K_{ AB,\xi}=0 \cr
(\sqrt{-K})_{,\bar\xi\bar\xi} + 2({\it ln}\phi)_{,\bar\xi}
(\sqrt{-K})_{,\bar\xi} + {1\over{4}}{1\over{\sqrt{-K}} }{K^{AB}}_{,\bar\xi}
K_{AB,\bar\xi}=0\cr }$$
\noindent
(c)$$ 8({\it ln}\phi)_{,\xi\bar\xi}+K^{-1}{K^{AB}}_{,\xi} K_{AB,\bar\xi}
 = 0\eqno(2.26)$$
\indent
The complete set of these conditions, complemented by the basic requirement
$ -K=- det(K_{AB})>0,$ is sufficient to assure $R_{\mu\nu}=0,$ but we do
not claim that all of these conditions are (independently)
necessary.\hfil\break
\indent
To end this section we rewrite equations (2.26) in matrix notation.  Let
$(\gamma )_{AB}=K_{AB}$ be the 2x2 symmetric matrix corresponding to
the Killingian part of the metric (1.1).  Observe that ${K_A}^R=
\gamma\epsilon$ and $K^{AB}=-\rho^2\gamma^{-1}$ with $\epsilon=
\Vert \epsilon_{AB}\Vert$,
then equations (2.26) can be cast into the form\hfil\break
$$a)\ \ \ (\rho\gamma^{-1}\gamma_{,\xi})_{,\bar\xi}+(\rho\gamma^{-1}
\gamma_{,\bar\xi})_{,\xi}=0$$
$$ b) \cases{ ({\it ln } \phi^{-2})_{,\xi} =
                            {{ {({\tenit ln}\rho)_{,\xi\xi}}\over
                            {({\tenit ln}\rho)_{,\xi}}} + {1\over{4({\tenit ln}
                                 \rho)_{,\xi}}} tr(\gamma_{,\xi}
                                 \gamma^{-1})^{2}}\cr
          ({\it ln}\phi^{-2})_{,\bar\xi} =
                             {({\tenit ln}\rho)_{,\bar\xi\bar\xi}
                            \over{({\tenit ln}\rho)_{,\bar\xi}}} +
                                       {1\over{4({\tenit ln}\rho)_{,\bar\xi}}}
                                  tr(\gamma_{,\bar\xi}\gamma^{-1})^{2} \cr } $$
$$ c)\ \ \ ({\it ln}\phi^{-2})_{,\xi{\bar\xi}}=
               ({\it ln}\rho)_{,\xi}({\it ln}\rho)_{,\bar\xi}+
                {1\over{4}}tr(\gamma^{-1}_{,\xi}
              \gamma_{,{\bar\xi}})\eqno(2.27)$$
\noindent
where we have set $K=-\rho^{2}.$ Equation (2,27c) is a consistency equation
of (2,27b)(see also ref. [9]).\hfil\break
\hfil\break
\noindent
${ \bf 3. The \ Potential \ Space. }$\hfil\break
\indent
The Ernst potential [13] ${\cal E}$ has proved to be a very useful tool for
finding exact solutions [14]. To define the potential ${\cal E}$, we need the
existence of a time-like Killing vector $Z_{\mu}$.  Then one finds
that the Einstein equations in vacuum\hfil\break
$$R_{\mu\nu}Z^{\nu}=0\eqno(3.1)$$
\noindent
implies that\hfil\break
$${Q^{\mu\alpha}}_{;\alpha}={\tilde Z}^{\mu ;\alpha}_{\ \ \ \ ;\alpha}=0
\Leftrightarrow Z^{\alpha}Q_{[\alpha\mu ;\nu]}=0\eqno(3.2)$$
\noindent
where $Q_{\mu \nu}={\tilde Z}_{\mu ;\nu} =Z_{\mu ;\nu}+{1\over{2}}
\epsilon_{\mu\nu\alpha\beta}Z^{\alpha ;\beta}.$  Furthermore one finds that
the Lie derivative of $Q_{\mu\nu}$ with respect to ${\bf Z}$ vanishes.  This
allows us to define the Ernst potential
${\cal E}_{\mu}=Z^{\alpha}Q_{\alpha\mu}$,
whose integrability conditions are just (3.2) and the vanishing
of the Lie derivative of
$Q_{\mu\nu},i.e.$\hfil\break
$$2{\cal E},_{[\mu ;\nu]}=Z^{\alpha}Q_{[\alpha\mu ;\nu]}-
{\it L}_{{\bf Z}}Q_{\mu\nu}=0\eqno(3.3)$$
\indent
The Einstein field equations in vacuum in terms of the Ernst potential
read [3],[14]\hfil\break
$$( {\cal E}+ \bar{\cal E} )\rho^{-1}(\rho{\cal E}_{,\mu})^{;\mu}=
2{\cal E}_{,\mu}\bar{\cal E}^{,\mu}.\eqno(3.4)$$
$$and\ \ \ \ \ \ {\cal E} \rightarrow \bar{\cal E}$$
\noindent
These equations can be derived from the Lagrangian\hfil\break
$${\cal L} = \rho{ 1\over{2f^{2}} } {\cal E}_{,\mu}\bar{\cal E }^{,\mu} \ \ \
f=Re{\cal E} \eqno(3.5)$$
\indent
Generation techniques consist in finding invariant transformations of the
Lagrangian (3.5). This is equivalent to finding the isometry group of the
metric\hfil\break
$$ dS^{2}={ 1\over{2f^{2}} }d{\cal E}d{\bar{\cal E}}.\eqno(3.6) $$
\indent
This isometry group is $SU(1,1)$ [14] which is isomorphic to $SL(2,{\bf R}).$ A
straightforward calculation shows that the metric (3.6) of the potential
space (defined by this metric) can be cast into the form [3]\hfil\break
$$dS^{2}={1\over{4}}tr(dG dG^{-1})\eqno(3.7)$$
\noindent
where the 2x2 matrix G is given by \hfil\break
$$ G = {-1\over{{\cal E}+\bar{\cal E}}}
\pmatrix{ 1+{\cal E}{\bar{\cal E}} & 1-{\cal E}{\bar{\cal E}}+{\cal E}
          -{\bar{\cal E}}\cr
           -1+{\cal E}{\bar{\cal E}}+{\cal E}-{\bar{\cal E}}& -1-{\cal E}
            {\bar{\cal E}}\cr}.$$
$$\eqno(3.8) $$
\indent
The matrix (3.8) is an element of the group $SU(1,1)$ restricted to $G^{2}=1.$
An other parametrization of (3.7) belongs to the group $SL(2,{\bf R})$, it
reads [6],[10]\hfil\break
$$G={1\over{f}}\pmatrix{f^{2}+\epsilon^{2}&-\epsilon\cr
                         -\epsilon&1\cr} $$
$$ \eqno(3.9) $$
\noindent
where ${\cal E}=f+i\epsilon.$  Observe that $det G=1$ in both cases, but $G$ in
(3.9) is symmetric. Now we return to the
field equations in the potential space.  It is clear that the field equation
(3.4) can be derived from the Lagrangian\hfil\break
$${\cal L}={1\over{4}}\rho tr(G_{,M}G^{-1,M}).\eqno(3.10)$$
\indent
We are interested on fields depending on two coordinates $X^1$ and $X^2$ .
Let $z=X^1 +iX^2$ be the complex variable which $G$ depends on.
Equation (3.4) transforms into [6],[10]\hfil\break
$$(\rho G_{,z}G^{-1})_{,\bar z}+(\rho G_{,\bar z}G^{-1})_{,z}=0\eqno(3.11)$$
\noindent
derived from Lagrangian (3.10).  They are the chiral equations with
$G\epsilon SL(2,{\bf R})$ and $G=G^{T}$.\hfil\break
\hfil\break
\noindent
${\bf 4.  The\ SL(2,{\bf R})-Chiral\ Equations}$\hfil\break
\indent
It is now clear why we are interested in developing a technique for solving
chiral equations with the group $SL(2,{\bf R}).$  Solitonic methods for
doing so are given in ref.[15].
We want to give an another method using an ansatz
due to Neugebauer and Kramer [8]. Let us outline it [7].\hfil\break
\indent
Suppose $G$ is an element of a Lie group $H$ which
depends on a set of parameters $\lambda^{a};\ a=1,...,p$ with
$\lambda^{a}=\lambda^{a}(z,{\bar z})$ such that these parameters form
minimal surfaces on a Riemannian space
$V_{p}\ ({\it HM},\lambda^a:{\it M}^4\rightarrow V_p)$:\hfil\break
$$(\rho\lambda^{a}_{,z})_{,\bar z}+(\rho\lambda^{a}_{,\bar z})_{,z}+
2\rho\gamma^a_{\ bc}\lambda^{b}_{,z}\lambda^{c}_{,{\bar z}}=0\eqno(4.1)$$
$$ \ \ \ \ \ \ \ \ \ \ \ \ a, b, c=1,...,p,$$
\noindent
where $ \gamma^a_{\ bc}$ are the Christoffel symbols of $V_{p}$.  In terms
of these parameters the chiral equation (3.11) reads\hfil\break
\noindent
a)$$A_{i;j}+A_{j;i}=0$$
\noindent
b)$$A_{i;j}=-[A_{i},A_{j}]\eqno(4.2)$$
\noindent
where $A_{i}=(\partial_{\lambda^i}G)G^{-1}$. $A_i$ is the Maurer-Cartan
form of the
group $H$ and therefor it belongs to the corresponding Lie algebra
$\cal G$ of $H$.  Equation (4.2a) is the Killing equation on $V_{p}$
for each component of matrix $A_{i}.$  Thus we write $A_{i}$ in spinor-like
notation\hfil\break
$$A_{i}=\xi^{j}_{i}\sigma_{j},\eqno(4.3)$$
\noindent
where $\{\xi_{i}\}$ is a linearly independent set of Killing vectors on
$V_{p}$ and $\{\sigma_{j}\}$ is a base of the vector space $\cal G$. Let
$L_{a}$ be the left action of $H_c$ on $H$,
being $H_c$ the group of
matrices on $H$ which does not depend on $z$
and ${\bar z},$ i.e. $C\epsilon H_c$
means $C_{,z}=C_{,{\bar z}}=0.$  Then the equivalence relation:
$A^{c}_{i}\sim A_{i}$ iff there exist $C\epsilon H_c$ such that
$A^{a}_{i}=A_{i} o L_{a},$ separates the set $\{A_{i}\}$ into equivalence
classes.  Let $TB=\{A_{i}\}/\sim$ be a set of representatives of each class.
Map each representative into the group building the set $B:=\{G\vert G=exp \
A_{i}, A_{i}\epsilon TB\}.$  If we know the set $B$, we can obtain the whole
set $H$ by left action of $H_c$ on $H$.
We apply it to the group
$SL(2,{\bf R}).$  Let first $p=1$ be the one dimensional Riemannian space
$V_{1}.$  All one dimensional Riemannian spaces are flat and the geodesic
equation (4.1) reduces to\hfil\break
$$(\rho\lambda_{,z})_{,{\bar z}}
+(\rho\lambda_{,{\bar z}})_{,z}=0,\eqno(4.4)$$
\noindent
which is the Laplace equation in $z,{\bar z}$ coordinates.  The Killing
equation (4.2a) now reads\hfil\break
$${A_{1}}_{,\lambda}=0,\ \ \ \ A_{1}=(\partial_{\lambda}G)G^{-1}.\eqno(4.5)$$
\noindent
That means that $A_{1}$ is a constant matrix of
${\cal G}=sl(2,{\bf R}).$ We need
then the matrices $G$ to be symmetric. Hence it is convenient to choose the
left action\hfil\break
$$L_{c}(G)=CGC^{T}\eqno(4.6)$$
\noindent
in order to have $L_{c}(G)$ also symmetric.  With this left action the
equivalence relation reads\hfil\break
$$A^{c}_{1}=A_{1}oL_{c}=CA_{1}C^{-1}.\eqno(4.7)$$
\noindent
The representatives of the classes can be the
2x2, traceless and real Jordan matrices.  There is only one representative,
i.e.\hfil\break
$$TB=\{A_{1}\}/\sim =\{ \pmatrix{0&1\cr
                            \alpha&0\cr} \}.$$
$$\eqno(4.8)$$
\noindent
Now we map $A_{1}$ into the group $H$, using its definition
$A_{1}=(\partial_{\lambda}G)G^{-1},$ i.e., solving this matrix differential
equation. We have\hfil\break
$$G_{,\lambda}=A_{1}G,\eqno(4.9)$$
\noindent
and get\hfil\break
$$G_{11,\lambda}=G_{12} \ \ \ \ \ \ G_{12,\lambda}=G_{22}$$
$$G_{12,\lambda}=\alpha G_{11} \ \ \ \ G_{22,\lambda}=\alpha G_{12}$$
\noindent
which imply the differential equations\hfil\break
$$ G_{11,\lambda\lambda } - \alpha G_{11}= 0 \ \ \ , G_{12,\lambda\lambda}
-\alpha G_{12}=0.$$
\noindent
We separate these differential equations in three cases, $\alpha=a^{2}>0$,
$\alpha =-a^{2}<0$ and $\alpha = 0$.
The results are given in table 1.\hfil\break\indent
We suppose now $p=2$ and $V_{2}$ a 2-dimensional Riemannian space.  But all
$V_{2}$ Riemannian spaces are conformally flat and therefore have a
metric which may  be written in the form\hfil\break
$$ dS^{2}={ {d\lambda d\tau}\over{(1+k\lambda\tau)^{2}}} .\eqno(4.10)$$
\noindent
In reference [2] and [8] it was shown that the $V_{p}$
spaces must be symmetric
(i.e. all covariant derivatives of the Riemannian tensor vanish), this
implies that $k$ must be a constant.
Now we choose a base of the Killing vector
space on $V_{2}.$  Let this base be\hfil\break
$$\xi_{1}={{1}\over{2V^{2}}}(k\tau^{2}+1,k\lambda^{2}+1)$$
$$\xi_{2}={{1}\over{V^{2}}}(-\tau,\lambda) \ \ \ \ \
V=(1+k\lambda\tau)$$
$$\xi_{3}={{1}\over{2V^{2}}}(k\tau^{2}-1,1-k\lambda^{2}).\eqno(4.11)$$
\noindent
With this set of Killing vectors the commutation relations (4.2b)
read\hfil\break
$$[\sigma_{1},\sigma_{2}]=-4k\sigma_{3}$$
$$[\sigma_{2},\sigma_{3}]=4k\sigma_{1}$$
$$[\sigma_{3},\sigma_{1}]=-4\sigma_{2}.\eqno(4.12)$$
\noindent
We have to put $k=-1$ in order to have the commutation relations of
${\it sl}(2,{\bf R}).$  A representation of ${\it sl}(2,{\bf R})$
is\hfil\break
$$\sigma_{1}=2\pmatrix{-1&0\cr
                       0&1\cr},
  \sigma_{2}=2\pmatrix{0&b\cr
                       a&0\cr},
  \sigma_{3}=2\pmatrix{0&-b\cr
                       a&0\cr}\ ,\ \ ab=1\ ,$$
$$\eqno(4.13a)$$
\noindent
which is a base of the vector space ${\cal G}={\it sl}(2,{\bf R}).$  Now
we map the algebra into the group.  From the definitions
$A_{\lambda}=(\partial_{\lambda}G)G^{-1}=\xi^{j}_{1}\sigma_{j}$ and
$A_{\tau}=(\partial\tau G)G^{-1}=\xi^{j}_{2}\sigma_{j}$ we obtain
 [9]\hfil\break
$$G={{1}\over{1-\lambda\tau}}\pmatrix{c(1-\lambda)(1-\tau)&e(\tau-\lambda)\cr
                                       e(\tau -\lambda)&d(1+\lambda)
                                       (1+\tau )\cr}$$
$$-e^{2}=cd \ \ , a={e\over{c}} \ \ , b=-{e\over{d}}.\eqno(4.13b)$$
\noindent
where $\lambda$ and $\tau$ must fulfill the geodesic equation (4.1).
With the metric (4.10) we get\hfil\break
$$(\rho\lambda_{,z})_{,{\bar z}}+(\rho\lambda_{,{\bar z}})_{,z}+
{{4\tau}\over{1-\lambda\tau}}\rho\lambda_{,z}\lambda_{,{\bar z}}
=0,\eqno(4.14)$$
\noindent
and the other one by changing $\lambda$ for $\tau$ and viceversa.
The other possibility is to put $k=0$ in (4.12), but this case
corresponds to the one-dimensional subalgebras of ${\it sl}(2,{\bf R})$
studied before with $\lambda \rightarrow \lambda + \tau$ . The
corresponding equation (4.1) will be the Laplace equation for
$\lambda$ and $\tau$ separately.  Observe
that $det \ G=1$ if we put $d=-1, c=1$ in (4.13) and $det\ G=-1$ if we put
$d=1=c.$\hfil\break
\indent
Let us introduce a real potential $\alpha$ by\hfil\break
$$\alpha_{,z}=-{1\over{2}}\rho tr(G_{,z}G^{-1}_{,z}).\eqno(4.15)$$,

\noindent
and a similar equation for $\alpha_{,\bar z}$, with ${\bar z}$ in plase of $z$.
Since the chiral equations imply $\alpha_{,z\bar z} = \alpha_{, \bar z z}$,
the integrability condition of $\alpha$ follows
from the chiral equation (3.11) [8].  This potential can be calculated
separately for each case.  For the one dimensional subspaces we have\hfil\break

$$\eqalign{
\alpha_{,z}&= {1\over{2}}\rho\ tr (G_{,z}G^{-1})^2\cr
           &= {1\over{2}}\rho\ tr(G_{,\lambda} G^{-1})^2 (\lambda_{,z})^2 \cr
           &= {1\over{2}}\rho\ tr A^{2}(\lambda_{,z})^{2}\cr},\eqno(4.16)$$
\noindent
and for the two-dimensional subspaces, a straightforward calculation
gives\hfil\break
$$\alpha_{,z}=-2\rho{{(\lambda -\tau)^{2}}\over{(1-\lambda\tau)^{4}}}
\lambda_{,z}\tau_{,z}.\eqno(4.17)$$
\noindent
These potentials will be important in the following section.\hfil\break
\hfil\break
\noindent
${\bf 5. The\ Unified\ Point\ of\ View}$\hfil\break
\hfil\break
\indent
In this section we apply the technique developed in section 4 to the
results of sections 2 and 3.\hfil\break
\indent
Equations (2.27a) and (3.11) are chiral equations, but there is a difference
between them.  In the first case $det \gamma =-\rho^{2},$ in the second
one $det \ G=1.$  We can transform equation (2.27a) in order to have a
(3.11)-like equation.  First observe that
$ \sqrt{-K}= \rho ={1\over{2}}(\xi +{\bar\xi}).$ is a solution of equation
(2.17).
Then we can write
$\xi =X^1 +iX^2 =z=\rho + i \zeta.$  Now define a matrix G such
that $\gamma =\rho G$
with $det \ G=-1.$ It is easy to see that $G$ fulfills just the chiral
equation (3.11).  Let us parametrize the matrix $G$ in Papapetrou form
$dS^{2}=e^{-2u}(e^{2k}dzd{\bar z}+\rho^{2}d\psi^{2})-e^{2u}
(dt+\omega d\psi)^{2},i.e.$ \hfil\break
$$G=-{{f}\over{\rho}}\pmatrix{-{1\over{f^2}}\rho^2+\omega^{2}&\omega\cr
                                \omega&1\cr}=
                                {1\over F}
                     \pmatrix{-F^{2}+\omega^{2}&\omega\cr
                                \omega&1\cr},
                               \ \ \  f=e^{2u},$$
$$\eqno(5.1)$$
\noindent
where $F=-\rho /f.$  Compare (5.1) with (3.9).
That means the following: if we have a
solution of the chiral equations in the potential space, it is also a
solution of the chiral equation in the spacetime.  This transformation
was given first in ref. [16] (see also ref. [17]).  We study the
subspaces on the spacetime.  There are the three cases of table 1 and
(4.13b).  For the first case on table 1 we choose $4bca^{2}=-1$ to
obtain\hfil\break
$$\gamma =
\pmatrix{ \rho (be^{a\lambda} -{1\over{4ba^{2}}}e^{-a\lambda})
        &\rho(bae^{a\lambda}+{{1}\over{4ba}}e^{-a\lambda})\cr
        &                        &                        \cr
          \rho(bae^{a\lambda}+{1\over{4ba}}e^{-a\lambda})
        &\rho(ba^{2}e^{a\lambda}-{1\over{4b}}e^{-a\lambda})\cr}.$$
$$\eqno(5.2)$$
\noindent
In order to obtain the function $\phi$ in (2.1), compare (2.27b) with
(4.15)\hfil\break
$$({\it ln}\rho^{1\over{2}}\phi^{-2})_{,z}=
a^2\rho(\lambda_{,z})^{2}\eqno(5.3).$$
\noindent
So, for each solution of the Laplace equation (4.4) we have a solution
of the Einstein's equations in vacuum.  It is easy to see that\hfil\break
$$\rho^{2} e^{-4u}=\rho^{2}{1\over{f^{2}}}=\omega^{2}-{1\over{a^{2}}},
\eqno(5.4)$$
\noindent
where we have used the Papapetrou parametrization (5.1).  This means that
(5.2) belongs to the Weyl's class (see ref. [14] eq. (18.22)) or to the van
Stockum class [18] by $a\rightarrow \infty $.
We can write the solution (5.2) in a
better-known form.  Take a (4.6) transformation with\hfil\break
$$C=\pmatrix{ C_{1} &C_{2}\cr
            aC_{4}  &C_{4}\cr}\eqno(5.5)$$
\noindent
we find that\hfil\break
$$\gamma =\pmatrix{ b\rho(C_{1}+aC_{2})^{2}e^{a\lambda}+c\rho
                   (C_{1}-aC_{2})^{2}e^{-a\lambda}&2\rho ba(C_{1}
                   +aC_{2})C_{4}e^{a\lambda}\cr
            2\rho ba(C_{1}+aC_{2})C_{4}e^{a\lambda} & 4ba^{2}C^{2}_{4}
             \rho e^{a\lambda} \cr},$$
$$\eqno(5.6)$$
\noindent
then which in the Papapetrou form reads $2u=a\lambda +{\it ln}\rho ,
4ba^{2}C^{2}_{4}=-1,\omega =$ constant.\hfil\break
\indent
An other example on this subspace is the solution of the Laplace equation
$\lambda = { n\over{a}}{\it ln}\rho .$ If we substitute this $\lambda$
in (5.2) then we find a cylindrically symmetric solution of the Einstein's
equations with one restriction (equation (20.7) in ref. [14] with
$B^{2}a_{1}a_{2}=-1/4.$ See also ref. [9]). Now we apply the transformation
(4.6) to get\hfil\break
$$CGC^{T}=A\pmatrix{C^{2}_{+}&C_{+}d_{+}\cr
                   C_{+}d_{+}&d^{2}_{+}\cr}
                   e^{a\lambda}+B
           \pmatrix{C^{2}_{-}&C_{-}d_{-}\cr
                   C_{-}d_{-}&d^{2}_{-}\cr}
                   e^{-a\lambda}$$
$$\eqno(5.7)$$
\noindent
which is the general cylindrically symmetric solution of the Einstein's
equations in vacuum, (equation (20.7) in ref. [14] with $c=C_{-}/d_{-}$,
$a_{1}=bd^{2}_{+}, a_{2}=-{{1}\over{4ba^{2}}}d^{2}_{-}).$\hfil\break
\indent
The second case in table 1 corresponds to the solution\hfil\break
$$\gamma =\pmatrix{-{\rho\over{a}}cos(a\lambda +\psi_0 )&\rho
                      sin(a\lambda + \psi_0)\cr
                       \rho sin(a\lambda +\psi_0)&
                      a\rho cos(a\lambda +\psi_0)\cr }$$
$$\eqno(5.8)$$
\noindent
where $b^{2}a^{2}=1,$ in order to have $det G =-1$.  The function $\phi$ is
then a solution of the differential equation\hfil\break
$$({\it ln}\rho^{1/2}\phi^{-2})_{,z}=-a^2\rho(\lambda_{,z})^{2}.\eqno(5.9)$$
\noindent
It is easy to see that\hfil\break
$$\rho^{2}e^{-4u}=\omega^{2}+1,\eqno(5.10)$$
\noindent
which corresponds to the Lewis' class (see eq. (18.22) in ref. [14]).  The
third case of table 1 corresponds to the degenerated class $f=0$
[8].\hfil\break
\indent
There is only one case in the two-dimensional subsapces.  In the spacetime one
gets the solution\hfil\break
$$\gamma ={{1}\over{(1-\lambda\tau )}}
\pmatrix{\rho (1-\lambda)(1-\tau)&\rho(\tau -\lambda)\cr
         \rho(\tau -\lambda)&-\rho (1+\lambda)(1+\tau )\cr }.$$
$$\eqno(5.11)$$
\noindent
Comparing (2.27b) with (4.17) we can obtain the function $\phi$ of (2.1)
in terms of the parameters $\lambda$ and $\tau$ , we arrive at\hfil\break
$$({\it ln}\rho\phi^{-2})_{,z}=-2\rho{{(\lambda -\tau)^{2}}\over
                                 {(1-\lambda\tau)^{4}}}\lambda_{,z}\tau_{,z}
\eqno(5.12)$$
\noindent
where $\lambda$ and $\tau$ are solutions of equation (4.14).\hfil\break
\indent
Now we deal with the subspaces in the potential space.  Here we need that
$det G =1$, but this can only happen in the first case in table 1.  We
obtain\hfil\break
$$G=\pmatrix{ be^{a\lambda}+{1\over{4ba^{2}}}e^{-a\lambda}&bae^{a\lambda}-
            {1\over{4ba}}e^{-a\lambda}\cr
            &             &           \cr
            bae^{a\lambda}-{1\over{4ba}}e^{-a\lambda}&
            ba^{2}e^{a\lambda}+{1\over{4b}}e^{-a\lambda}\cr}.$$
$$\eqno(5.13)$$
\noindent
If we compare (5.13) with (3.9), we find that\hfil\break
$$f={{1}\over{ba^{2}e^{a\lambda}+{1\over{4b}}e^{-a\lambda}}},
\epsilon =-{{bae^{a\lambda}-{1\over{4ab}}e^{-a\lambda}}\over
            {ba^{2}e^{a\lambda}+{1\over{4b}}e^{-a\lambda}}}.\eqno(5.14)$$
\noindent
If we set $b=1/2, a=e^{\psi_0}$, the Ernst potential ${\cal E}=f+i\epsilon$
can be written as\hfil\break
$${\cal E}={1\over{a}}
[sech (a\lambda +\psi_0)-i \ tangh(a\lambda +\psi_0)]\eqno(5.15)$$
\noindent
(compare it with eq. (17.32) of ref. [14]).  The function $k$ in the
Papapetrou metric (5.1) can be obtained by integrating
the equation [16]\hfil\break
$$ k_{,z}=\sqrt{-2}\rho{{\cal E}_{,z}\bar{\cal E}_{,z}
\over{4f^2}}.\eqno(5.16)$$
\noindent
We find that $k={\sqrt{2}\over4}\alpha$ being $\alpha$ potential (4.16).
It is easy to see that\hfil\break
$$f^{2}=-\epsilon^{2}+1/a^{2},\eqno(5.17)$$
\noindent
which means that those solutions belong to the Papapetrou's class (see
eq. (18.16) of ref. [14]).\hfil\break
\indent
Finally for the two-dimensional subspaces (4.13b) in the potential
space we must have $det G=1$.  This can be done if we write
$\zeta =\lambda ={\bar\tau}$, i.e.\hfil\break
$$G={{1}\over{1-\zeta{\bar\zeta}}}
     \pmatrix{(1-\zeta)(1-{\bar\zeta})&-i(\zeta -{\bar\zeta})\cr
     -i(\zeta-{\bar\zeta})&(1+\zeta)(1+{\bar\zeta})\cr}.\eqno(5.18)$$
\noindent
A direct comparation of (5.18) with (3.9) shows that ${\cal E}=(1-\zeta)/
(1+\zeta)$. The geodesic equation (4.14) reads\hfil\break
$$(\zeta{\bar\zeta}-1){1\over{\rho}}[(\rho\zeta_{,z})_{,\bar z}+
(\rho\zeta_{,\bar z})_{,z}]=4{\bar\zeta}\zeta_{,z}\zeta_{,\bar z}\eqno(5.19)$$
\noindent
(compare this equation with eq. (18.28) of ref. [14]).  These solutions
belong to the Tomimatsu-Sato class.  The Kerr solution in terms of
the $\zeta$ potential is $\zeta =px-iqy\ \ (p^{2}+q^{2}=1)$
in prolate coordinates $x,y$ [14], [19].
The function $k$ in front of the Papapetrou metric can be
calculated.  From (4.15) we again arrive at
$k={\sqrt{2}\over4}\alpha$, but now
$\alpha$ is the potential (4.17).  All the results are shown in table
2.\hfil\break
\hfil\break
\noindent
{\bf 6. Conclusions}\hfil\break
\hfil\break
\indent
We have shown that the solutions of the Einstein's equations with two
commuting Killing vectors can be separated naturally in equivalent
classes given by the subgroups of $SL(2,{\bf R}).$  Those classes
coincide with the classes of solutions, the one parametric subgroups of
$SL(2,{\bf R})$ give the Weyl's class, van Stockum class, the general
cylindrically symmetric solutions, the Lewis' and the degenerated classes
in the spacetime, and the Papapetrou's class in the potential space.
The group $SL(2,{\bf R})$ itself gives the Tomimatsu-Sato class in the
potential space and a corresponding class in the space time.  The left
action of the group $SL(2,{\bf R})_C$ over $SL(2,{\bf R})$ is just the
invariance transformations of Neugebauer and Kramer [16], i.e. they are the
Ehlers and the Buchdahl transformations to obtain new solutions from a
seed one.  Then we have shown that a classification of these solutions
can be done by the classification of the subgroups of the invariance
group.\hfil\break
\vskip 2cm

\noindent
{\bf Acknowledgements}\hfil\break
\indent
The authors would like to thank the conversations with Riccardo
Capovilla, and the referee's remarks to this paper. This work is
supported in part by CONACYT- Mexico
\hfil\break
\noindent
{\bf References}\hfil\break
[1] Matzner, R. A., and Misner, C.W.  Phys. Rev.$\underline {154}$ (1967),
1229.\hfil\break
\indent
Misner, C.W.  Phys. Rev. $\underline {D18}$ (1978), 4510.\hfil\break
[2] S\'anchez, N. Phys. Rev. $\underline {D26}$ (1982), 2589.\hfil\break
[3] Neugebauer, G. and Kramer D. In $``$ Galaxies, axisymmetric
systems\hfil\break
\indent
and relativity$''$ Ed. M.A. H. MacCallum (1985).\hfil\break
[4] Belinsky, V.A.  and Zakharov, V.E. Zh. Eksp. Teor. Fis.
$\underline {75}$ (1978), 1953.\hfil\break
[5] Whitman, A.P. and Stoeger, W.R. Gen. Rel. Grav. $\underline {24},$
(1992), 641\hfil\break
[6] See for example Matos, T. Rev. Mex.Fis. $\underline {36}$ (1990), 340
or D\'\i az, C.M.\hfil\break
\indent
Rev. Mex. Fis. $\underline {34}$ (1988), 1.\hfil\break
[7] Matos, T., Rodr\'\i guez, G. and Becerril, R. J. Math. Phys.
$\underline {33}$, (1992), 3521.\hfil\break
[8] Neugebauer, G. and Kramer, D. "Stationary Axisymmetric Einstein Maxwell
\hfil\break
\indent
Fields Generated by B\"acklund Transformations".
Jena Preprint 1989.\hfil\break
[9] Matos, T. Rev. Mex. Fis.$\underline {35}$ , (1989), 208.\hfil\break
[10] Matos, T. Phys. Lett . $\underline {A131}$ , (1989), 423.\hfil\break
[11] Clement, G. Gen. Rel. Grav. $\underline {18}$ (1987), 5.\hfil\break
[12] Matos, T. Ann. Phys. (Leipzig) $\underline {46}$ , (1989),
462.\hfil\break
[13] Ernst, F.J.V. Phys. Rev. $\underline {167}$ (1968), 1175.\hfil\break
\indent Neugebauer, G. and Kramer, D. Ann. Phys. (Leipzig)
$\underline {24}$ (1969), 62.\hfil\break
\indent Ernst, F. and Pleba\'nski, J. Ann. Phys.$\underline{107}$ (1977),
266\hfil\break
[14] Kramer, D., Stephani, H., MacCallum, M. and Herlt, E.\hfil\break
\indent
"Exact Solutions of Einstein's Field Equations". VEB-DVW, 1980.\hfil\break
[15] Kramer, D., Neugebauer, G. and Matos, T. J. Math. Phys.
$\underline {32}$\hfil\break
\indent
(1991), 2727. Mendoza, A. and Restuccia, A. J. Math. Phys.\hfil\break
\indent
$\underline {32}$ (1991), 480.\hfil\break
[16] Neugebauer, G. and Kramer, D. in ref. [13].\hfil\break
[17] Nakamura, Y. J. Math. Phys. $\underline {24}$ (1983), 606.\hfil\break
[18] van Stockum, W.J.  Proc. Roy. Soc. Edinburgh
$\underline {A57}$ (1937), 135.\hfil\break
[19] Tomimatsu, A. and Sato, H. Prog. Theor. Phys.
$\underline {50}$ , (1973), 95.\hfil\break
\vfil\eject
\vskip 2cm
\hoffset 0cm
\vskip 3cm
\nopagenumbers
\vbox{\tabskip=0pt \offinterlineskip
\def\tablerule{\noalign{\hrule}}
\halign to435pt  {\strut#&\vrule#\tabskip=1em plus5em&
\hfil#& \vrule#& \hfil #\hfil & \vrule #& \hfil #\hfil & \vrule #&
\hfil#& \vrule#& \hfil #\hfil & \vrule #&
\hfil #& \vrule # \tabskip=0pt \cr \tablerule
&&\omit \hidewidth A \hidewidth &&\omit\hidewidth $\alpha$
\hidewidth &&\omit\hidewidth G \hidewidth &&\omit\hidewidth det G
\hidewidth &&\omit\hidewidth tr$A^2$ \hidewidth &\cr\tablerule
&&      1.       &&                         &&   &&    &&    &\cr
&&  $\pmatrix{ 0 &1\cr \alpha & 0} $        &&  $ a^2$     &&
$G=b\pmatrix{ 1 & a\cr a & a^2 }$ $e^{a\lambda}$  + c
$\pmatrix{ 1 & -a \cr -a & a^2}$ $e^{-a\lambda}$ &&  $4 bca^2$ && $2a^2$
&\cr
&&          && 	             &&             &&      &&  &\cr\tablerule
&&    2.    &&               &&                   &&      &&  &\cr
&&          && $ -a^2$ &&
$G = b \pmatrix{ cos(a\lambda + \psi_0)  & - a\ sin(a\lambda + \psi_0)\cr
          -a\ sin(a\lambda+\psi_0)        & -a^2cos(a\lambda +\psi_0)}$
      && $ -b^2 a^2 $      && $ -2a^2 $   &\cr
&&          &&               &&                   &&      &&  &\cr\tablerule
&&    3.    &&             &&         &&          &&          &\cr
&&	    && 	 0         && $G=\pmatrix{ b\lambda+c
& b \cr  b & 0 \cr}$    && $ -b^2$     &&  0   &\cr
&&          &&             &&          &&          &&         &\cr\tablerule
\noalign{\smallskip} }}
\centerline{ Table 1. One-dimensional subspaces of
 SL(2,{\bf R}). $a,\ b,\ c,\ \psi_0$, constants.  }
\vskip 4cm
\vbox{\tabskip=0pt \offinterlineskip
\def\tablerule{\noalign{\hrule}}
\halign to400pt  {\strut#&\vrule#\tabskip=1em plus5em&
\hfil#& \vrule#& \hfil #\hfil & \vrule #& \hfil #\hfil & \vrule #&
\hfil#& \vrule#& \hfil #\hfil & \vrule #&
\hfil #& \vrule # \tabskip=0pt \cr
\tablerule
&&\omit \hidewidth Subspace \hidewidth &&\omit\hidewidth Spacetime
\hidewidth &&\omit\hidewidth  Potential space\hidewidth
&\cr\tablerule
&&         && 1. Weyl's class && Papapetrou's class & \cr
&&	    && Von Stockum class
 	                    &&   &\cr
&&          && (with $a\rightharpoonup\infty )$ 	 && & \cr
&&          &&              &&                     & \cr
&&          && Cylindrically symmetric           &&                         &
\cr
&& One-dimensional && solutions (with $\lambda =	{n\over a}ln \rho$)&&     &
\cr
&&	    && 	                                 &&                         & \cr
&&	    && 2. Lewis' class                   && $\exists{\kern -2mm /}$ &\cr
&&          && 3. Degenerated class $f=0$        && $\exists{\kern -2mm /}$
&\cr
\tablerule
&&Two-dimensional  && 2-dim. class &&Tomimatsu-Sato class	            &\cr
&&          &&                                  &&                          &
\cr
&&          &&                                  &&                          &
\cr\tablerule
\noalign{\smallskip} }}
\centerline{ Table 2. Classification of the solutions}

\end